\documentclass[prd,twocolumn,notitlepage,nofootinbib,longbibliography,superscriptaddress]{revtex4-1}

\usepackage[english]{babel}
\usepackage{amsmath,amssymb}
\usepackage{graphicx,float,placeins}
\usepackage{comment}
\usepackage{diagbox}
\usepackage{booktabs}
\usepackage[colorlinks=true, allcolors=blue]{hyperref}
\usepackage{textgreek}

\newcommand{\figh}{4.75cm}

\begin{document}

\title{Photon Absorption in a Doubly Special Relativity Model with Undeformed Free Propagation and Total Momentum Conservation}

\author{J. M. Carmona}
\email{jcarmona@unizar.es}
\affiliation{Departamento de Física Teórica and Centro de Astropartículas y Física de Altas Energías (CAPA), Universidad de Zaragoza, Zaragoza 50009, Spain}

\author{J. L. Cortés}
\email{cortes@unizar.es}
\affiliation{Departamento de Física Teórica and Centro de Astropartículas y Física de Altas Energías (CAPA), Universidad de Zaragoza, Zaragoza 50009, Spain}

\author{F. Rescic}
\email{filip.rescic@uniri.hr}
\affiliation{Departamento de Física Teórica and Centro de Astropartículas y Física de Altas Energías (CAPA), Universidad de Zaragoza, Zaragoza 50009, Spain}
\affiliation{University of Rijeka, Faculty of Physics, Rijeka 51000, Croatia}

\author{M. A. Reyes}
\email{mkreyes@unizar.es}
\affiliation{Departamento de Física Teórica and Centro de Astropartículas y Física de Altas Energías (CAPA), Universidad de Zaragoza, Zaragoza 50009, Spain}

\author{T. Terzi\'c}
\email{tterzic@phy.uniri.hr}
\affiliation{University of Rijeka, Faculty of Physics, Rijeka 51000, Croatia}

\begin{abstract}
The lack of a dynamical framework within doubly special relativity theories has impeded the development of a corresponding phenomenology of modified interactions. In this work we show that in a model based on the classical basis of $\kappa$-Poincaré and total momentum conservation, one has a well-defined cross section of the photon-photon annihilation process, once a prescription for the channel treatment is set. The modification of the interaction can lead to observable effects in the opacity of the Universe to very high-energy gamma rays when the gamma-ray energy approaches the energy scale of the deformation. The magnitude and observability of this deformation are examined as functions of the gamma-ray energy and source distance.
\end{abstract}

\maketitle

\section{Introduction}
\label{sec:intro}

The propagation of high-energy astrophysical gamma rays is impeded by interactions with the low-energy photon backgrounds present in the Universe. These interactions result in a decrease in the gamma-ray flux, with some photons being converted into electron-positron pairs~\cite{nikishov1962,Gould:1967zza}. For gamma-ray energies between $10^{10}$--$10^{14}$\,eV, interactions with the extragalactic background light (EBL) play the leading role. Instead, between $10^{14}$--$10^{19}$\,eV, the interaction with the cosmic microwave background (CMB) is dominant. For energies above $10^{19}$\,eV, the radio background (RB) becomes the relevant background~\cite{DeAngelis:2013jna}.

Thanks to the advancement of gamma-ray astronomy in the last decades, one can use the gamma-ray transparency of the Universe as a window to scenarios that go beyond special relativity (SR). At high energies, this is particularly motivated by quantum gravity models~\cite{Addazi:2021xuf}, for which the combination of high energies and astrophysical distances serve as amplifiers of these possible effects of new physics.

The most studied scenario involves the violation of the Lorentz symmetry of SR (Lorentz invariance violation or LIV). This can be done by including high-dimensional Lorentz violating operators at the Lagrangian level in the effective field theory framework, suppressed by a power $n$ of a new physics scale $\Lambda$, usually related with the Planck energy ($E_\text{Pl}\approx 1.22 \times 10^{28}\,\mathrm{eV}$). Considering LIV effects in the photon sector leads to a modification of the usual photon dispersion relation, $E=|\vec{p}|$, with an effective energy dependent squared mass $\pm E^{(2+n)}/\Lambda^n$ (see~\cite{Carmona:2024thn} and references therein), where $E$ is the energy of the photon, $n$ denotes the order of correction (being the most commonly studied scenarios the linear, $n=1$, and quadratic, $n=2$, cases), and the sign of the effective mass defines the subluminal ($-$) or superluminal ($+$) scenarios.

Modifying the photon energy-momentum relation can lead to several observable consequences. For instance, in the linear case, the absence of observed birefringence effects constrains $\Lambda$ to be many orders of magnitude above the Planck scale~\cite{Gotz:2014vza}. In the quadratic superluminal case, anomalous photon decays, like photon splitting or vacuum pair production, set the strongest bounds, which reach three orders of magnitude below the Plank energy~\cite{LHAASO:2021opi}. These decays are not allowed in the quadratic subluminal case, but other anomalous interactions, like the photon-photon pair production, and anomalous time-of-flight (time delays), put bounds around eight orders of magnitude below the Plank scale~\cite{Abdalla:2019krx}.

Considering that the highest energy of detected gamma rays is around the petaelectronvolt ($10^{15}$\,eV)~\cite{LHAASO:2021gok}, and that in fact present analyses use the significantly higher flux at TeV ($10^{12}$\,eV) energies, it may seem surprising that the observed gamma-ray flux suppression has been able to impose constraints on the LIV scale $\Lambda$ of the order of the Planck energy or higher in the linear case, and of order $10^{-8}\, E_\text{Pl}$ in the quadratic case~\cite{Terzic:2021rlx}, since for these values of the scale of new physics $(E/\Lambda)^n \ll 1$. This fact can be understood by noticing that LIV strongly modifies the usual threshold of pair production of special relativity, with an additional $\pm E^{(2+n)}/(4m_e^2 \Lambda^n)$ term~\cite{Jacob:2008gj}, where $m_e$ is the electron mass.

A less explored scenario beyond SR is doubly special relativity (DSR)~\cite{AmelinoCamelia:2008aez}. In this framework there is not a privileged system of reference, in contrast to the LIV case, but transformations between inertial observers are $\Lambda$-deformed versions of the usual Lorentz transformations that guarantee the presence of a relativity principle in the theory~\cite{Amelino-Camelia:2000stu}. DSR kinematics is more difficult to analyze than LIV kinematics because, besides a possible deformation of the dispersion relation, in the case of DSR, the conservation laws involved in the kinematics of particle interactions  are necessarily changed with respect to those of SR to preserve the relativity principle.

As a consequence of the link established by the relativity principle between the dispersion relation and the energy-momentum composition law, the modifications brought about by DSR kinematics are generally milder than in the LIV case. For instance, the relativity principle forbids emergent particle instability, so the stringent constraints from the appearance of anomalous decays in the LIV case cannot be applied. We find an additional example in the photon-photon pair production, whose modified threshold condition, as shown in Section~\ref{sec:threshold}, is modified by a factor $(1+E/\Lambda)$, which makes the correction phenomenologically irrelevant if $E/\Lambda \ll 1$.

For these reasons, the phenomenological study of modified interactions in the DSR scenario has been usually disregarded in favour of anomalous time-of-flight analyses, which typically produce constraints on the scale of new physics of the order of the Planck scale. However, it has been shown that anomalous time delays are not a necessary consequence of DSR models~\cite{Carmona:2022pro}. Then, a scale of new physics several orders of magnitude below the Planck scale opens the possibility of observable effects in the transparency of the Universe at energies available in current observatories and detectors. In any case, such a scale must be greater than the energy scales probed in laboratory and accelerator experiments, at least of the order of $10^{12}\,$eV, since otherwise, the effects of new physics would have already been observed~\cite{Carmona:2023luz}.

The aim of this article is to investigate how photon absorption changes in this kind of DSR scenario. In Section~\ref{sec:DSR} we show the specific DSR model used in this work. Using the aforementioned model, we will compute the deformation of the threshold and cross section of the photon-photon pair production process. The details of the calculation can be found in Sections~\ref{sec:threshold} and \ref{sec:cross_section}. In Section~\ref{sec:transparency} we show how the calculated deformation with respect to SR can lead to observable effects in the survival probability of the photons at the detector. Finally, in Section~\ref{sec:conclusions} we discuss the results and expectations for current and future detectors.

\section{A DSR model with undeformed free particle propagation and total momentum conservation}
\label{sec:DSR}

The development of DSR models was initially inspired by quantum group theory, manifesting as a $\kappa$-parametric deformation of the Poincaré algebra, commonly referred to as $\kappa$-Poincaré~\cite{Majid:1994cy,Amelino-Camelia:2000stu,KowalskiGlikman:2002we}. Within this algebraic framework, the bicrossproduct basis of $\kappa$-Poincaré has been one of the most extensively examined DSR models~\cite{Kowalski-Glikman:2004fsz}. However, different bases of the $\kappa$-Poincaré Hopf algebra lead to different phenomenological implications. 

A DSR model that includes a modification of the dispersion relation generally predicts photon time delays that would be noticeable at scales $\Lambda$ comparable to the Planck energy~\cite{Amelino-Camelia:2011ebd}. However, in a model which does not alter the Poincaré algebra at the single-particle level, like the \emph{classical basis} of $\kappa$-Poincaré, the Casimir (and therefore the dispersion relation) and the one-particle Lorentz transformations remain identical to those of SR~\cite{KowalskiGlikman:2002we}. This implies that constraints arising from time delays do not affect this particular DSR model~\cite{Carmona:2017oit,Carmona:2022pro}, allowing to consider lower values of the scale of new physics, until the deformation at the level of the composition law starts playing a role in particle interactions.

As outlined in the introduction, our interest lies in the possibility of deformation scales that are significantly lower than the Planck scale. Therefore, a DSR model grounded in the classical basis of $\kappa$-Poincaré is particularly well-suited for the current phenomenological study. A different DSR model, in which time delays for massless particles are also absent, was employed in a previous work exploring the implications of DSR on the transparency of the Universe~\cite{Carmona:2021lxr}. Nevertheless, the classical basis has been identified as a robust framework for addressing consistency issues in DSR models~\cite{Carmona:2023luz}, making it the natural choice for the analysis presented in this study.

In a $\kappa$-Poincaré-based DSR model, compatibility between a deformed composition of momenta in a two-particle system and relativistic invariance can be achieved by identifying this composition with the coproduct of the generators of translations in the corresponding $\kappa$-Poincaré Hopf algebra model~\cite{Carmona:2016obd}. Then, the simplest scenario for exploring DSR effects in interactions is to assume that the deformed relations among the momenta of interacting particles due to invariance under translations can be expressed as an equality between the total momentum of the initial and final states. While more general scenarios, due to the noncommutativity of momentum composition, could be considered\footnote{These more general scenarios will be addressed in a future work.}, we focus on this simplified scenario in the present study.

The total momentum of a system of two particles of four-momenta $a$ and $b$ is given in a DSR scenario based on the classical basis of $\kappa$-Poincaré by the following composition law~\cite{Borowiec2010}: 
\begin{align}
    (a\oplus b)_0&=a_0 \Pi(b)+\frac{1}{\Pi(a)}\left(b_0+\frac{\Vec{a}\cdot\Vec{b}}{\Lambda}\right), \label{MCL0} \\
    (a\oplus b)_i&=a_i \Pi(b)+b_i\,,
    \label{MCLi}
\end{align}
where 
\begin{equation}
    \Pi(a)=\frac{a_0}{\Lambda}+\sqrt{1+\frac{a_0^2-|\Vec{a}|^2}{\Lambda^2}}\,.
\end{equation}
Let us denote the four-momenta of the initial high-energy and soft photons as $k$ and $q$, respectively. Similarly, the four-momenta of the produced electron and positron can be denoted by $p_{-}$ and $p_{+}$. Due to the non-commutativity of the composition of momenta, we can define four different conservation laws,
\begin{align}
    k\oplus q &= p_{-} \oplus p_{+}\,, \label{ch1}\\
    k\oplus q &= p_{+} \oplus p_{-}\,, \label{ch2}\\
    q\oplus k &= p_{-} \oplus p_{+}\,, \label{ch3}\\
    q\oplus k &= p_{+} \oplus p_{-}\,. \label{ch4}
\end{align}
The previous equations provide four alternative channels through which the interaction can proceed. In the following sections, we will demonstrate that it is possible to derive an unambiguous prediction for the cross section of each channel. In the absence of a theoretical framework incorporating DSR kinematics to dictate how the different channels should be combined, we will consider two alternative scenarios: (1) assume that all channels are equally probable and use the average of the results for the different channels; (2) assume that the theory selects the channel that minimizes the cross section (minimum interaction principle).

\section{Modified threshold of the process}
\label{sec:threshold}

Let us consider the production of an electron and positron with energy-momentum $p_-=(E_-,\vec{p}_-)$ and $p_+=(E_+,\vec{p}_+)$,  by the electromagnetic interaction of a high-energy photon, of energy-momentum $k=(E,\vec{k})$, with a low-energy photon of the electromagnetic cosmic background, of energy-momentum $q=(\omega,\vec{q})$,
\begin{equation}\label{pairc}
    \gamma(k) + \gamma_\text{soft}(q) \to e^-(p_-) + e^+(p_+).
\end{equation}

In SR, the threshold condition for this reaction to occur can be obtained by imposing the conservation of the squared total four-momentum of the system before and after the scattering, and identifying its minimum value,
\begin{equation}
    (k+q)^2 \,=\, (p_-+p_+)^2 \geq 4 m_e^2\,.
    \label{eq:genthreshold}
\end{equation}
where the left part of the equality can be evaluated in the laboratory reference frame, giving $(k+q)^2=2E\omega(1-\cos\theta)$, while the right part of the equality can be evaluated in the center of mass reference frame of the electron and positron, from which one obtains $(p_-+p_+)^2 = 4 (m_e^2 + |\vec{p}_+|^2) \geq 4 m_e^2$. Then, the SR threshold condition can be stated as 
\begin{equation}
    \frac{2E\omega(1-\cos\theta)}{4m_e^2} \geq 1 \,.
    \label{eq:SR_th}
\end{equation}

We can repeat the same argument for each of the DSR channels shown in Eqs.~\eqref{ch1}-\eqref{ch4}. The identification of the composition of momenta with the coproduct of the generators of translations in the Hopf algebra implies that the total momentum of a system of two particles transforms as the momentum of a particle. Then, in the model based on the classical basis, the total momentum transforms linearly and its square is a relativistic invariant. For illustrative purposes we will show the details of the procedure for the first channel, Eq.~\eqref{ch1}. The other channels can be solved analogously.

Considering the first channel, one gets for $(k\oplus q)^2$ in the laboratory frame,
\begin{align}
    (k\oplus q)^2&= \frac{2E\omega(1-\cos\theta)}{1+E/\Lambda} \left( 1+ \frac{1}{2} \frac{E\omega(1-\cos\theta)}{\Lambda^2(1+E/\Lambda)} \right) \notag \\ & \approx \frac{2E\omega(1-\cos\theta)}{1+E/\Lambda}\,,
    \label{s-initial}
\end{align}
where in the last step we have made use of the fact that the soft photon energy $\omega$ is much smaller than the high-energy scale $\Lambda$.

To get the modified threshold condition, one should compare the previous expression with the minimum value of the invariant, which can be obtained using the center of mass reference frame of the final electron and positron. One can generalize the definition of this reference frame to DSR as $(p_{-}\oplus p_{+})_i=0$ for every spatial component $i=1,2,3$. From this definition, one gets that $(p_{-})_i=-(p_{+})_i/\Pi(p_{+})$ in the classical basis. Combining this relation and the dispersion relation of the electron and positron (which are not modified with respect to SR), one can write the four-momentum of the emitted pair as \mbox{$p_{-}=(\sqrt{m_e^2+|\vec{p}_{+}|^2/\Pi^2(p_{+})},-\vec{p}_{+}/\Pi(p_{+}))$} and \mbox{$p_{+}=(\sqrt{m_e^2+|\vec{p}_{+}|^2},\vec{p}_{+})$}. This way, one can express the value of the invariant in the center of mass reference frame as a function of $p_+$, 
\begin{align}
    (p_{-}\oplus p_{+})^2&=\Bigg[\sqrt{m_e^2+\frac{|\vec{p}_{+}|^2}{\Pi^2(p_{+})}}\Pi(p_{+}) \notag \\ &+\frac{1}{\Pi(p_{-}(p_+))}\left(\sqrt{m_e^2+|\vec{p}_{+}|^2}-\frac{|\vec{p}_{+}|^2}{\Pi(p_{+})\Lambda}\right)\Bigg]^2\,.
    \label{s-final}
\end{align}

\begin{figure}[tp]
    \centering
    \includegraphics[height=\figh]{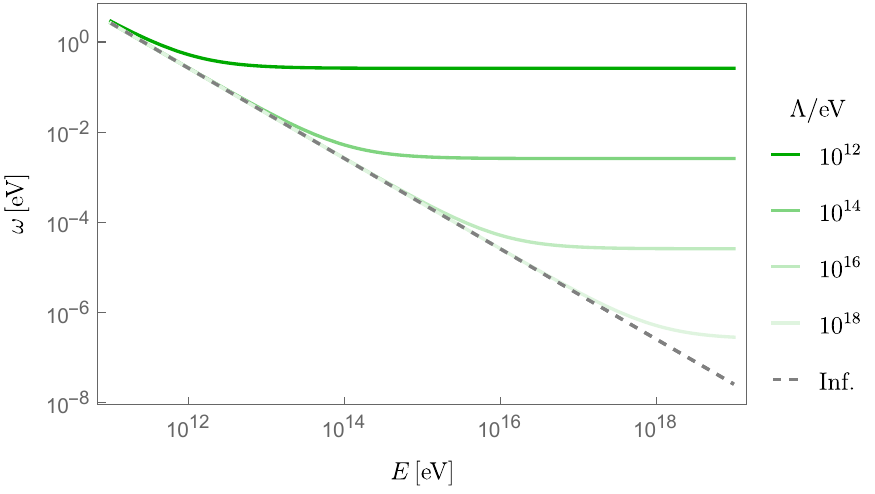}
    \caption{In green solid lines, values of initial photon energies that solve the threshold equality in Eq.~\eqref{eq:DSR_th} for $\theta=\pi$ and different values of the scale $\Lambda$. The kinematically allowed region in each case corresponds to the values of $E$ and $\omega$ that lie above each solid line.}
    \label{fig:threshold}
\end{figure}

After using the mass shell condition, $p_+^2 = m_e^2$, one can check that the previous expression is a monotonically increasing function of the single variable $|\vec{p}_{+}|$, so that the minimum value will be at $|\vec{p}_{+}|=0$,
\begin{equation}
    (p_{-}\oplus p_{+})^2 \Big|_{|\vec{p}_{+}|=0} =4 m_e^2\left(1+\frac{m_e^2}{\Lambda^2} \right)\approx 4 m_e^2\,.
    \label{s-min}
\end{equation}
Then, using Eqs.~\eqref{s-initial} and \eqref{s-min}, we obtain that the DSR modified threshold condition for the first channel can be stated as
\begin{equation}
    \frac{2E\omega (1-\cos\theta)}{4m_e^2 (1+E/\Lambda)}\gtrapprox 1 \,. \label{eq:DSR_th}
\end{equation}
The kinematically allowed region of $E$ and $\omega$ for the interaction to occur through this first channel is shown in Fig.~\ref{fig:threshold} for different values of the scale $\Lambda$.

An exchange in the ordering between the electron and positron does not modify the results; then, the obtained threshold is also valid for the second channel, Eq.~\eqref{ch2}. In contrast, for the third and fourth channels, Eqs.~\eqref{ch3} and~\eqref{ch4} respectively, one can check that
\begin{align}
    (q\oplus k)^2&= \frac{2\omega E(1-\cos\theta)}{1+\omega/\Lambda} \left( 1+ \frac{1}{2} \frac{\omega E(1-\cos\theta)}{\Lambda^2(1+\omega/\Lambda)} \right) \notag \\ & \approx 2E\omega(1-\cos\theta)\,,
\end{align}
where in the last step we used that $\omega\ll\Lambda$. Therefore, the threshold condition for the third and fourth channels, Eqs.~\eqref{ch3} and~\eqref{ch4} respectively, is approximately the same as in SR, i.e., Eq.~\eqref{eq:SR_th}.

\section{Modified cross section of the process}
\label{sec:cross_section}

The cross section of a process is proportional to the integral of the squared transition amplitude of that process over the final states phase space. Let us introduce the notation $\mathcal{F}$ to refer to such dimensionless integral,
\begin{equation}
    \mathcal{F}=\int [\mathcal{DPS}] \, |\mathcal{M}|^2 \quad\rightarrow\quad \sigma=\frac{1}{\mathcal{K}}\, \times \mathcal{F}\,,
    \label{eq:F}
\end{equation}
where we have introduced the notation $[\mathcal{DPS}]$ for the integral measure over the final particles phase space, $\mathcal{M}$ for the matrix element, and $1/\mathcal{K}$ for the proportionality factor, which only depends on the initial free states.

In this section, we are going to study the value of the function $\mathcal{F}$ and the prefactor $1/\mathcal{K}$ for the annihilation of two photons producing an electron-positron pair. In the case of SR, it is convenient to define a dimensionless version of the SR invariant, 
\begin{equation}
    \bar s \doteq \frac{(k+q)^2}{4m_e^2} = \frac{2E\omega(1-\cos\theta)}{4m_e^2}\,,
    \label{eq:bar_s}
\end{equation}
which reduces the SR threshold condition to $\bar s \geq 1$. The inverse of the SR initial state factor becomes $\mathcal{K}_\text{SR}=8m_e^2 \bar s$, and the function $\mathcal{F}$ can be written as
\begin{align}
    {\cal F}_\text{SR} &\equiv \mathcal{F}_\text{BW}(\bar s) \doteq 4\pi \alpha^2 \Bigg[ \left(2 + \frac{2}{\bar s}-\frac{1}{\bar{s\,}^2} \right) \notag \\ &\times \ln \left(\frac{1+\sqrt{1-1/\bar s}}{1-\sqrt{1-1/\bar s}}\right) - \left(2+\frac{2}{\bar s}\right) \sqrt{1-1/\bar s} \Bigg] \,.
    \label{eq:BW_F}
\end{align}
Eq.~\eqref{eq:BW_F} is the well-known result computed by Breit and Wheeler in~\cite{Breit:1934zz}, to which we will refer from now on as the Breit-Wheeler formula.

In the case of LIV, one still has a well-defined effective field theory framework, the Standard Model Extension~\cite{Colladay:1998fq,Kostelecky:2008ts}, to compute the modified matrix element and the corresponding function $\mathcal{F}_\text{LIV}$. 
Unfortunately, most authors settle on approximating the value of the cross section using effective approaches. In~\cite{Carmona:2024thn}, a new expression for the function $\mathcal{F}_\text{LIV}$ was provided based on an explicit calculation, overcoming the limitations of previous approximate approaches. 
One could also consider an LIV modification on the kinematical prefactor $1/{\cal K}$ from the initial state, but such modification will be an irrelevant correction proportional to the ratio $E/\Lambda$ and can then be neglected. 

In the DSR scenario, there is not yet a well-defined dynamical framework for calculating the matrix element and the corresponding function $\mathcal{F}_\text{DSR}$. However, one can get an unambiguous prediction of the function $\mathcal{F}$ for each of the channels of the model presented in Sec.~\ref{sec:DSR}. For illustrative purposes, we will show the details of the procedure for the first channel, Eq.~\eqref{ch1}, but the other channels can be solved analogously.

We will refer to the integral in Eq.~\eqref{eq:F}, when applied to the first channel (Eq.~\eqref{ch1}), as ${\mathcal{F}_{\text{DSR},1}}$. The invariance of the matrix element and the phase-space measure under the deformed Lorentz transformations implies that the dimensionless function ${\mathcal{F}_{\text{DSR},1}}$ can only depend on dimensionless ratios of invariant quantities. Given that there are three available invariant quantities, $(k\oplus q)^2$, $m_e^2$ and $\Lambda^2$, the function $\mathcal{F}_{\text{DSR},1}$ can only depend on two independent ratios,
\begin{equation}
        \mathcal{F}_{\text{DSR},1} \equiv f\left(\frac{(k\oplus q)^2}{4m_e^2},\frac{(k\oplus q)^2}{\Lambda^2}\right)\,.
        \label{eq:f}
\end{equation}
The first argument of the function $f$ provides a generalization of the SR invariant for the first channel. Then, it is convenient to introduce the notation
\begin{equation}
    \bar \tau_{1}\doteq \frac{(k\oplus q)^2}{4m_e^2}\approx \frac{2E\omega(1-\cos\theta)}{4m_e^2(1+E/\Lambda)}\,,
    \label{eq:tau_1}
\end{equation}
analogously to the case of SR, Eq.~\eqref{eq:bar_s}. The threshold condition for the first channel now reads $\bar\tau_1 \geq 1$. Taking into account that we are studying values of $\Lambda\sim E$ and that $E\gg m_e\gg \omega$, then one can approximate
\begin{equation}
\mathcal{F}_{\text{DSR},1} \approxeq f(\bar{\tau}_1,0)\,.
\end{equation}
For consistency with the SR limit, one has
\begin{equation}
    \lim_{(E/\Lambda) \to 0}\mathcal{F}_{\text{DSR},1}=\lim_{(E/\Lambda) \to 0} f(\bar{\tau}_1,0) = f(\bar{s},0)=\mathcal{F}_{\text{BW}}(\bar{s})\,.
\end{equation}
Therefore,
\begin{equation}
    \mathcal{F}_{\text{DSR},1}\approxeq f(\bar{\tau}_1,0)=\mathcal{F}_\text{BW}(\bar\tau_1).
\end{equation}

Since the one-particle sector is not modified in the DSR model in consideration, and consequently the free propagation of particles are described as in SR, then the normalization factor from the initial state (containing the normalization of free particle states and the definition of flux) will not be modified with respect to SR. Then, the cross section associated to the first channel is
\begin{equation}   
    \sigma_{\text{DSR},1}(E,\omega,\theta) \approx \frac{1}{\mathcal{K}_\text{SR}(E,\omega,\theta)}\, \times {\mathcal{F}_\text{BW}}\left(\bar\tau_1(E,\omega,\theta)\right)\,.
    \label{eq:cross_1}
\end{equation}

\begin{figure}[tp]
    \centering
    \includegraphics[height=\figh]{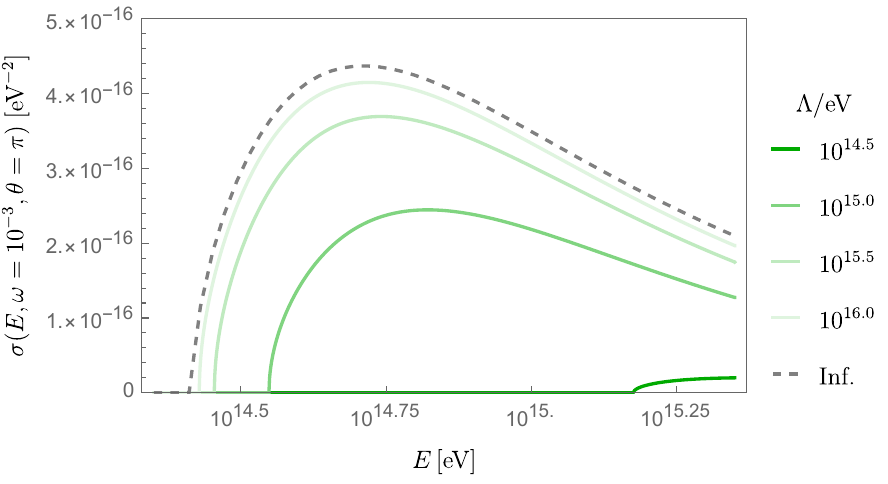}
    \caption{Cross section corresponding to the first two channels (Eqs.~\eqref{ch1} and~\eqref{ch2}) as a function of the gamma-ray energy, for $\omega=10^{-3}\,\mathrm{eV}$, for $\theta=\pi$, and for different values of the scale $\Lambda$. The cross section corresponding to the other two channels (Eqs.~\eqref{ch3} and~\eqref{ch4}) approximately coincides with the result of SR, i.e., the dashed curve.}
    \label{fig:cross_section}
\end{figure}

One can repeat the previous procedure for the other channels. Let us notice that for the second channel, Eq.~\eqref{ch2}, one finds the same invariant as in the first channel, i.e, $\bar \tau_2 =\bar \tau_1$. Then, following the same procedure, one arrives at the same modified cross section, Eq.~\eqref{eq:cross_1}. We show in Fig.~\ref{fig:cross_section} the modified cross section of these two channels for different values of the scale $\Lambda$. In contrast, for the third and fourth channels, the invariant is $\bar \tau_4 = \bar \tau_3 \approx \bar s$. Then, following the same procedure, one finds the same cross section as in SR.

Assuming the minimum interaction scenario, the transparency of the Universe to gamma rays will be controlled by the cross section, Eq.~\eqref{eq:cross_1} (channel with the smaller cross section). Instead, if we consider that all channels are equally probable, one should consider the average of the cross sections of all channels, i.e.,
\begin{equation}   
    \langle\sigma_\text{DSR,i}\rangle (E,\omega,\theta) \approxeq \frac{1}{\mathcal{K}_\text{SR}(E,\omega,\theta)}\times \langle{\mathcal{F}_\text{DSR,i}}\rangle\left(E,\omega,\theta\right) \,,
    \label{eq:cross_DSR}
\end{equation}
which in turn depends on the average of the function $\mathcal{F}$,
\begin{equation} \langle{\mathcal{F}_\text{DSR,i}}\rangle\left(E,\omega,\theta\right) = \frac{1}{4} \sum_i^4 {\mathcal{F}_\text{BW}}\left(\bar\tau_i(E,\omega,\theta)\right) \,.
\end{equation}
 
\section{Modified gamma-ray transparency of the Universe}
\label{sec:transparency}

The transparency of the Universe to gamma rays is usually characterized by the opacity $\tau(E,z_s)$, which is related with the photon survival probability according to
\begin{equation}
    \mathrm{Prob}(E,z_s) = \exp\left(-\tau(E,z_s)\right) \,,
    \label{eq:prob}
\end{equation}
where $E$ is the observed gamma-ray energy and $z_s$ is the redshift of the observed source. The opacity can be computed using the cross section of the photon absorption process, multiplied by the spectral density of background photons, and integrated over the angle between the momenta of the two photons producing the electron-positron pair and over the trajectory from the source to the detector:
\begin{align}
    \tau(E,z_s) \,=\, &\int_{0}^{z_s}dz\,\left|\frac{dl}{dz}\right|\int_{-1}^{1} d\cos\theta \left(\frac{1-\cos\theta}{2}\right) \notag \\ \times &\int_{\omega_\text{th}(E,\theta)}^\infty d\omega \; n(\omega,z) \,\sigma(E(1+z),\omega,\theta) \,,
    \label{eq:opa}
\end{align}
where $n$ is the spectral photon number density of the low-energy electromagnetic background, $\omega_\text{th}$ the minimum value of the energy of the soft photon necessary to satisfy the threshold condition, and $\sigma$ is either Eq.~\eqref{eq:cross_1} or Eq.~\eqref{eq:cross_DSR}, whether one assumes the minimum interaction scenario or equally probable channels, respectively. For the latter case, let us note that the use of Eq.~\eqref{eq:cross_DSR} is equivalent to compute the opacity for each channel and then averaging the result. In the following, we develop the calculation for the first channel, Eq.~\eqref{ch1}. Other channels can be solved analogously.

The first integral of Eq.~\eqref{eq:opa} contains the distance travelled by a photon per unit of redshift
\begin{equation}
    \frac{dl}{dz}\,=\,\frac{dl}{dt}\frac{dt}{dz} = \frac{-1}{(1+z)H(z)}\,,
    \label{eq:dldz}
\end{equation}
where $dl/dt=1$, and $H(z)$ is the redshift-dependent Hubble parameter. In the $\Lambda$CDM cosmological model
\begin{equation}
    H(z)= H_0\sqrt{\Omega_\text{m}(1+z)^3+\Omega_\Lambda}\equiv H_0 \,h(z)\,,
    \label{eq:lcdm}
\end{equation}
with $\Omega_\text{m}=0.3$ and $\Omega_\text{$\Lambda$}=0.7$ the matter and vacuum energy densities, respectively, and $H_0=70$ km s$^{-1}$ Mpc$^{-1}$ as the present value of the Hubble constant.\footnote{We used the fiducial value $H_0=70$ km s$^{-1}$ Mpc$^{-1}$ to avoid entering the discussion of the ``Hubble tension'', which is beyond the scope of this study. This is the same value assumed in~\cite{Saldana-Lopez:2020qzx} for the determination of the EBL model used in this work.}

We are going to restrict ourselves to energies below $10^{19}$\,eV, where the interactions with the RB can be neglected. Then, the photon density background under consideration contains contributions from the CMB and EBL,
\begin{equation}
    n(\omega,z)=n_{\text{CMB}}(\omega,z)+n_{\text{EBL}}(\omega,z)\,.
    \label{eq:n}
\end{equation}
The contribution from the CMB corresponds to the black-body spectrum
\begin{equation}\label{eq:ncmb}
    n_{\text{CMB}}(\omega,z) = \frac{(\omega/\pi)^2}{\exp(\omega/( 
    (1+z) T_0 \,k_\text{B}))-1} \,,
\end{equation}
where $T_0=2.73 \,\mathrm{K}$ is the present temperature of the CMB and $k_\text{B}$ is the Boltzmann constant. There is no general analytical formula for $n_{\text{EBL}}$ as a function of the redshift. Instead, one has to use a particular model, which is in turn based on a cosmological model. Here we use the EBL model of Saldana-Lopez et al. (2021)~\cite{Saldana-Lopez:2020qzx}, in which $\Lambda$CDM cosmology is assumed, with the values for the cosmological parameters given after Eq.~\eqref{eq:lcdm}.

Replacing Eqs.~\eqref{eq:dldz}, \eqref{eq:lcdm} and \eqref{eq:cross_1} in the calculation of the opacity (Eq.~\eqref{eq:opa}), and making the change of variables from $(\theta,\omega)$ to $(\bar\tau_1,\omega)$, one can rewrite the opacity corresponding to the first channel, $\tau_1(E,z_s)$, in the following compact form
\begin{align}
    \tau_\text{DSR,1}(E,z_s) \,=&\, \frac{m_e^2}{4 H_0 E^2} \int_0^{z_s} dz \;\frac{1+(1+z)E/\Lambda}{(1+z)^3 h(z)} \notag \\ &\times \int_{1}^{\infty} d\bar\tau_1\; \mathcal{F}_\text{BW}(\bar\tau_1) \int_{\omega_\text{th}(\bar\tau_1,z,E)}^{\infty} d\omega\; \frac{n(\omega)}{\omega^2} \,,
    \label{eq:opacity_1}
\end{align}
where now
\begin{equation}
    \omega_\text{th}(\bar\tau,z,E) = \frac{m_e^2 \,\bar\tau}{(1+z)E} \left(1+(1+z)\frac{E}{\Lambda}\right) \,.
\end{equation}

One can solve the other channels using the same procedure. Averaging those results, the opacity corresponding to the equally probable channels scenario can be obtained, 
\begin{equation}
    \langle\tau_\text{DSR,i}\rangle(E,z_s) \,=\, \frac{1}{4} \sum_i^4 \tau_\text{DSR,i} (E,z_s) \,.
    \label{eq:opacity_DSR}
\end{equation}

\section{Results and discussion}
\label{sec:discussion}

\begin{figure}[tbp]
    \centering
    \includegraphics[width=\linewidth]{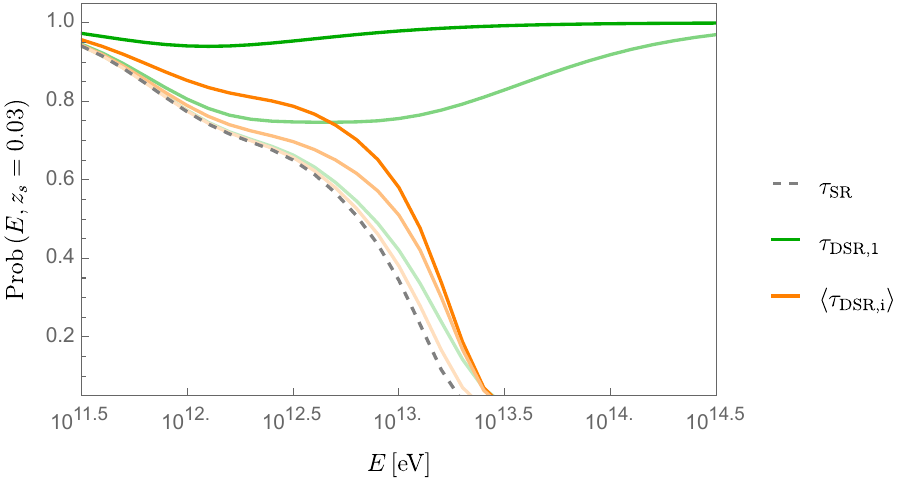}
    \includegraphics[width=\linewidth]{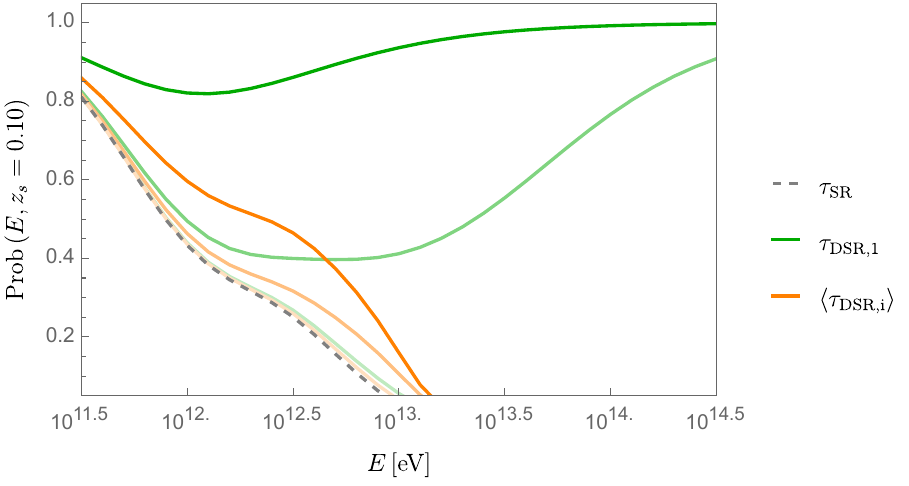}
    \caption{Survival probability for the minimum interaction (green) and equal probable channels (orange) scenarios, considering values of the scale $\Lambda/\mathrm{eV}=10^{12},10^{13}$ and $10^{14}$ (from darker to lighter), and two source redshifts, $z=0.03$ (upper plot) and $z=0.10$ (bottom plot). The grey dashed line represents the SR (Breit-Wheeler) case.}
    \label{fig:prob_z}
\end{figure}

As previously discussed, considering DSR anomalous interactions can decrease the transparency of the Universe to gamma rays of energies comparable to the value of the new physics scale $\Lambda$. 
For energies between $10^{10}$--$10^{14}\,$eV, opacity effects are only relevant for extragalactic sources, for which the EBL plays the dominant role.
To produce observable effects within this energy range, while ensuring compatibility with laboratory and accelerator experiments, values of $\Lambda$ between $10^{12}$--$10^{14}\,$eV should be used. In Fig.~\ref{fig:prob_z}, we show the result of the computation of the probability of survival (Eq.~\eqref{eq:prob}), as a function of the gamma-ray energy, for the minimal interaction and equally probable channels scenarios, and for two choices of extragalactic source redshifts.

\begin{figure}[tbp]
    \centering
    \includegraphics[width=\linewidth]{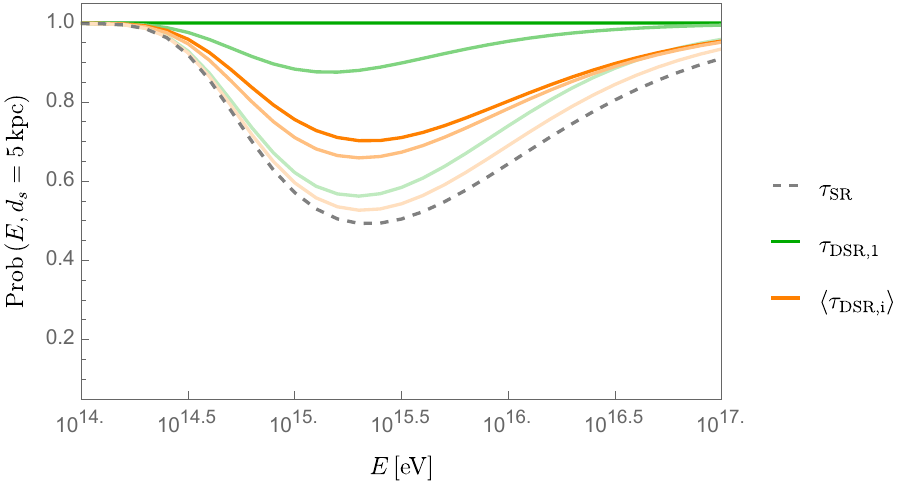}
    \includegraphics[width=\linewidth]{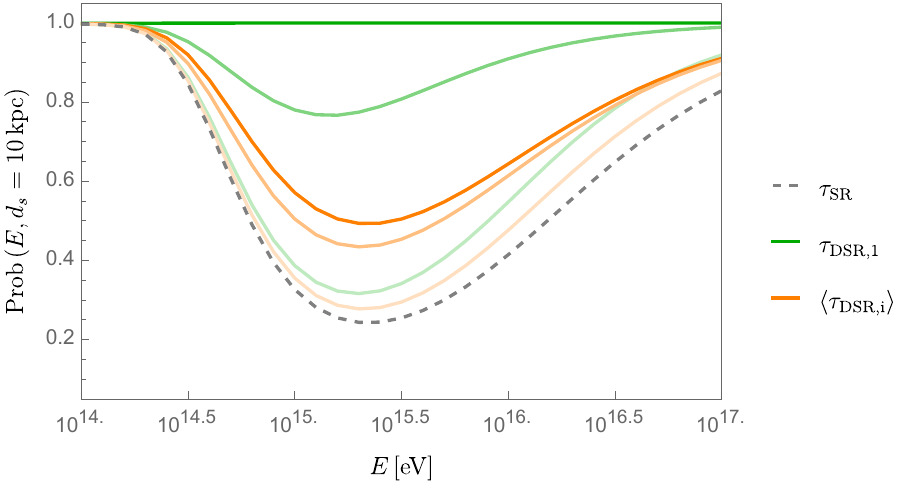}
    \caption{Survival probability for the minimum interaction (green) and equal probable channels (orange) scenarios, considering values of the scale $\Lambda/\mathrm{eV}=10^{14},10^{15}$ and $10^{16}$ (from darker to lighter), and two source distances, $5\,\mathrm{kpc}$ (upper plot) and $10\,\mathrm{kpc}$ (bottom plot). The grey dashed line represents the SR (Breit-Wheeler) case.}
    \label{fig:prob_d}
\end{figure}

On the other hand, galactic sources provide an opportunity to examine significant photon fluxes of energies of the order of the PeV and higher, for which the CMB is the relevant background.
For the local universe, it is a good approximation to disregard the redshift dependencies of the gamma-ray energy and the soft-photon background, so that the first integral in Eq.~\eqref{eq:opa} is just the Euclidean distance to the source, $d_s = \int_0^{z_s} |dl/dz|\, dz \approx z_s/H_0$. In Fig.~\ref{fig:prob_d}, we show the probability of survival for the two scenarios, for two choices of source distances within the galaxy, and for values of $\Lambda$ between $10^{14}$--$10^{16}\,$eV. Larger values of $\Lambda$ do not produce observable effects with respect to the SR case. For lower values of the scale of new physics, the probability of survival will swiftly converge to 1 in the minimum interaction scenario, and to the result of SR corresponding to half of the distance for the equally probable channels scenario.

Using Eqs.~\eqref{eq:opacity_1} and \eqref{eq:opacity_DSR}, one can extend the previous plots to a heat-map of the probability of survival as function of the gamma-ray energies and source redshift/distance. Fig.~\ref{fig:heatmap1} shows the one corresponding to the minimal interaction scenario, $\text{Prob}_{\text{DSR},1}(E,z_s)$, for different values of $\Lambda$. One can check that the first channel significantly affects the transparency of the universe to gamma rays, predicting full transparency for values of $\Lambda\sim10^{12}\,$eV, except at large distances and TeV energies. Instead, if equally probable channels are assumed, the averaging of the opacity softens to a large extent the differences with respect to SR at the level of the probability of survival.  

To better identify the range of energies and source redshifts for which deviations from SR are most significant, we present the difference in survival probabilities of DSR and SR, $\Delta\text{Prob}(E,z_s)$, in Fig.~\ref{fig:heatmap_delta}. The maximum and minimum differences are designated by the colours blue and red, respectively. This information can be used to determine the best source candidates for searching for photon flux anomalies. For instance, considering only the first channel (first column), for $\Lambda = 10^{14}$ eV, deviations from SR become noticeable close to 100 TeV for sources around $z_s = 10^{-3}$. A similar trend occurs for $\Lambda = 10^{12}$ eV, where deviations begin to appear at approximately 1 TeV for sources with redshifts around $z_s = 1$. In contrast to the equally probable channel (second column), where there is only a narrow strip of potential candidate sources, the minimum interaction scenario allows for a much broader range of potential sources to serve as probes for this effect.

\section{Conclusions}
\label{sec:conclusions}

We begin this section by discussing the differences between the results presented in this work and in our previous work~\cite{Carmona:2021lxr}, which was a first attempt to consider the effects of DSR in the propagation of high-energy astrophysical gamma rays. In~\cite{Carmona:2021lxr}, where the DSR model under consideration also deformed one-particle states, the modified kinematical prefactor from the initial state $1/\mathcal{K}$ was determined by the requirement that the cross section of each channel should remain invariant under the deformed Lorentz transformations that leave the squared total momentum of that channel invariant. This led to consider, for instance, for the first channel, $\mathcal{K}_\text{DSR,1} = 8 m_e^2 \bar{\tau}_1$. However, the use of Eq.~\eqref{eq:opa} for the computation of the photon opacity $\tau(E,z_s)$ is in correspondence to a convention\footnote{In SR, the cross section of a two-particle interaction of momenta $p_1,p_2$ and masses $m_1,m_2$ is conventionally defined with the factor $\mathcal{K}_\text{SR}=4\sqrt{(p_1\cdot p_2)-m_1^2 m_2^2}$ (see Eq.~(49.27) in~\cite{ParticleDataGroup:2024cfk}). For two photons, this corresponds to the value $\mathcal{K}_\text{SR}=8m_e^2 \bar{s}=4E\omega(1-\cos\theta)$ quoted in the text.} in the kinematical prefactor $\mathcal{K}$ appearing in the definition of the cross section $\sigma=\mathcal{F}/\mathcal{K}$ (Eq.~\eqref{eq:F}). Then, a modification of the kinematical prefactor requires a modification of Eq.~\eqref{eq:opa} accordingly. This, unfortunately, was not done in Ref.~\cite{Carmona:2021lxr}, leading to an inconsistency in the computation of the optical depth. Moreover, the adoption of a specific composition law, chosen primarily for its simple linear dependence on the four-momentum of each particle, introduced a certain degree of arbitrariness in the formulation of the DSR model. An additional limitation of that study was the assumption of equal probabilities for the different channels as the only possibility in the determination of the optical depth.

In contrast, in the present work we have used a DSR scenario defined by the classical basis of $\kappa$-Poincaré, which is physically motivated in order to solve consistency problems of general DSR models~\cite{Carmona:2023luz}. In this scenario, there is no modification of the kinematical prefactor, which follows then the same convention as in SR. Therefore, one can still use Eq.~\eqref{eq:opa} as the proper way to compute the optical depth. Let us note that, in this model, the cross section is not an invariant anymore. However, the relativistic structure of the theory is contained in the invariance of the function $\mathcal{F}$, which is proportional to the probability per unit time per unit volume of a single transition from an initial to a final state. Then, all the difference in the computation of the opacity in the SR, LIV and DSR models is in the use of different functions $\mathcal{F}$.

In case observations would show a higher transparency of the universe to gamma rays than what is expected, both LIV and DSR models could offer an explanation by modifying photon-photon interactions. In~\cite{Carmona:2024thn}, a computation of the gamma ray transparency of the Universe for a quadratic subluminal LIV model (and for various effective proposals for the photon-photon absorption cross section) was performed. That LIV model, likewise the DSR model studied in this work, predicts an increment of the gamma-ray probability of survival, providing an optimistic perspective for tests on current and future very- and ultra-high-energy gamma-ray observations. However, values of the scale of new physics close to the Planck energy are sufficient to produce observable effects of LIV on the universe transparency, in contrast to the DSR case, which requires much smaller values (of the order of the energy of the gamma ray). In the DSR case, the scale $\Lambda$ parametrizes the modified composition of momenta, leading to corrections depending on the ratio $E/\Lambda$ in the photon absorption. In the LIV scenario, the scale appears in a quadratic subluminal modification of the dispersion relation, and consequently the effect on photon absorption is controlled by the ratio $E^4/m_e^2\Lambda^2$. Then, the energy dependency of the probability of survival, and consequently the prediction for the detected flux, is also very different for each model, enabling the possibility to distinguish between them. A comparison of our current results of Fig.~\ref{fig:prob_z} and Fig.~\ref{fig:prob_d} with those of our previous study considering an LIV model~\cite{Carmona:2024thn}\footnote{The relevant comparisons are between the present Fig.~\ref{fig:prob_d} and the two plots in Fig.~6 of reference~\cite{Carmona:2024thn}, and between the present Fig.~\ref{fig:prob_z} and the bottom plots of Fig.~9 and Fig.~10 of reference~\cite{Carmona:2024thn}.} makes it apparent. 

Finally, it is also important to identify the limitations of the use of Fig.~\ref{fig:heatmap_delta} to identify source candidates to search for photon flux anomalies. Beyond identifying the optimal energy and redshift ranges where DSR effects are most pronounced in the survival probability, one should also take into account the astrophysical uncertainties and the experimental sensitivities at these energies. Moreover, the study of possible effects of new physics in the propagation of high energy photons should be also complemented by a study of the associated effects in the interactions involved in their production and detection. These considerations are out of the scope of the current paper, but will be addressed in a future work.

\section*{Acknowledgments}
This work is supported by the Spanish grants PGC2022-126078NB-C21, funded by MCIN/AEI/ 10.13039/501100011033 and `ERDF A way of making Europe’, grant E21\_23R funded by the Aragon Government and the European Union, and the NextGenerationEU Recovery and Resilience Program on `Astrofísica y Física de Altas Energías’ CEFCA-CAPA-ITAINNOVA, and by the Croatian Science Foundation (HrZZ) Project IP-2022-10-4595 and the University of Rijeka Project uniri-iskusni-prirod-23-24. The work of M.A.R. has been supported by the FPI grant PRE2019-089024, funded by MICIU/AEI/FSE. F.R. gratefully acknowledges the Erasmus Mobility program of the University of Rijeka. The authors would like to acknowledge the contribution of the COST Actions CA18108 ``Quantum gravity phenomenology in the multi-messenger approach'' and CA23130 ``Bridging high and low energies in search of quantum gravity''.

%

\end{document}